\documentstyle[prl,aps,twocolumn,psfig,floats]{revtex}

\begin{document}
\title{Theory of the density fluctuation spectrum of strongly correlated
  electrons}

\author{G. Khaliullin and P. Horsch}

\address{
Max-Planck-Institut f\"{u}r Festk\"{o}rperforschung,
Heisenbergstr.~1, D-70569 Stuttgart (Germany)}

\date{May 14, 1996}
\maketitle

\begin{abstract}
The density response function $N({\bf q},\omega)$ of the
two-dimensional $t$-$J$ model is studied starting from a mixed gauge
formulation of the slave boson approach. Our results for $N({\bf q}, 
\omega)$ are in remarkable agreement with exact diagonalization
studies, and provide a natural explanation of the anomalous features
in the density response in terms of the spin polaron nature of the
charge carriers.
In particular we have identified unexplained low energy structures in
the diagonalization data as arising from the coherent polaron motion
of holes in a spin liquid.

\noindent
PACS numbers: 71.27.+a, 71.45.Gm, 74.20.Mn
\end{abstract}
\pacs{PACS numbers: 71.27.+a, 71.45.Gm, 74.20.Mn}


Recent exact diagonalization studies\cite{toh95,ede95} 
of the dynamical density response $N({\bf q},\omega)$ at large 
momentum transfer have revealed several features unexpected from the
point of view of weakly correlated fermion systems: (i) a strong
suppression of low energy $2k_F$ scattering in the density response,
(ii) a broad incoherent peak whose shape is rather insensitve to hole
concentration and exchange interaction $J$, (iii) the very different
form of $N({\bf q},\omega)$ compared to the spin response function
$S({\bf q},\omega)$, which share common features in usual fermionic
systems.  
%

While considerable analytical work has been done to explain the spin
response of the $t$-$J$ model\cite{fuk} only few authors analysed $N({\bf q},
\omega)$. Wang {\it et al.}\cite{wan91} studied collective excitations in the
density channel and found sharp peaks at large momenta
corresponding to free bosons. Similar results were obtained by Gehlhoff and
Zeyher\cite{geh95} using the X-operator formalism. Lee et
al\cite{lee96} considered a model of bosons in a fluctuating gauge
field and found a broad incoherent density fluctuation spectrum at
finite temperature, due to the coupling of bosons to a quasistatic
disordered gauge field.

The aim of our paper is to show that the essential features observed
in the numerical studies can be obtained in the framework of the
Fermi-liquid phase of the $t$-$J$ model at zero temperature. Our main
findings are: (i) at low momenta the main effect of strong
correlations is to transfer spectral weight from particle-hole
excitations into a pronounced collective mode. Because of the strong
damping of this mode (linear in $q$) due to the coupling to the spinon
particle-hole continuum, this collective excitation is qualitatively
different from a sound mode. (ii) At large momenta we find a strict
similarity of $N({\bf q},\omega)$ with the spectral function of a
single hole moving in a uniform RVB spinon background. 
In this regime $N({\bf q},\omega)$ consists of a broad peak at high
energy whose origin is the fast, incoherent motion of bare holes. 
The polaronic nature of dressed holes leads to the formation of a
second peak at lower energy, which is more pronounced in $(\pi,0)$
direction in agreement with diagonalization studies\cite{toh95,ede95}.
For the static structure factor we find good agreement with numerical
results\cite{put94,gro94}.

Following Kotliar and Liu\cite{kot88} and Wang et al \cite{wan91} we
start from the $N$-component generalization of the slave-boson $t$-$J$
Hamiltonian, $H_{tJ}=H_t + H_J$:
\begin{eqnarray}
H_t &=& -\frac{2t}{N}\sum_{<i,j>\sigma} (f^+_{i\sigma} h^+_j
h_i f_{j\sigma} + h.c.),\nonumber  \\
H_J &=& \frac{J}{N}\sum_{<i,j>\sigma\sigma'} f^+_{i\sigma}
f_{i\sigma'}  f^+_{j\sigma'} f_{j\sigma},
\end{eqnarray}
where $\sigma =1,\cdots,N$ is the fermionic channel index, and 
$h_i$ denotes the bosonic holes. 
The number of auxiliary particles must obey the constraint  
$\sum_{\sigma} f^+_{i\sigma}f_{i\sigma} +h^+_ih_i=N/2$.
The original $t$-$J$ model is recovered for $N=2$.

The slave boson parametrization provides a straightforward description
of the strong suppression of density fluctuations of constrained
electrons through the representation of the density response in terms
of a dilute gas of bosons. A common treatment of model (1) is the
density-phase representation (``radial'' gauge\cite{rea83}) of the
bosonic operator $h_i=r_i \exp(i\theta_i)$ with the subsequent
$1/N$-expansion around the Fermi-liquid saddle point. While this gauge
is particularly useful to study the low energy and momentum
properties, it is not very convenient for the study of the density
response in the full $\omega$ and ${\bf q}$ space. Formally the latter
follows in the radial gauge from the fluctuations of $r_i^2$. If one
considers for example convolution type bubble diagrams, one realizes
that their contribution to the static structure factor is correctly of
order $1/N$, but is not proportional to the density 
of holes $\delta$ as it should be.
According to Arrigoni et al\cite{arr94} such unphysical
results originate from a large negative pole contribution in the 
$\langle r_{-\bf q}r_{\bf q}\rangle_{\omega}$ 
Green's function of the real field $r$, which is
hard to control by a perturbative treatment of phase fluctuations. 
We follow therefore Popov\cite{pop83} using the density-phase
treatment only for small momenta $q<q_0$, while keeping the original
particle-hole representation of  the density operator, $b^+b$, at
large momenta. More precisely $h_i=r_i\exp(i\theta_i) + b_i$, where
$b_i=\sum_{|{\bf q}|>q_0} h_{\bf q}\exp(i{\bf q}{\bf R}_i)$. The
cutoff $q_0$ is introduced dividing ``slow'' (collective) variables
represented by $r$ and $\theta$ from ``fast'' (single-particle)
degrees of freedom. As explained by Popov\cite{pop83} this ``mixed''
gauge is particularly useful for finite temperature studies to
control infrared divergences. We start formally with ``mixed'' gauge
and keep only terms of order $\delta$ and $1/N$ in the bosonic self
energies. In this approximation our zero temperature calculations
become quite straightforward: The cutoff $q_0<\delta$ actually does
not enter in the results and we arrive finally at the Bogoliubov
theory for a dilute gas of bosons moving in a fluctuating spinon
background. 

The Lagrangian corresponding to the model (1) is then given by
(the summation over $\sigma$ is implied)
\begin{eqnarray}
&L&=\sum_{i} \Bigl( {f^+_{i\sigma}}
  \bigl({{\partial}\over{\partial\tau}}-\mu_f\bigr) f_{i\sigma}
  +b^+_i \bigl( {{\partial}\over {\partial\tau}}-\mu_b\bigr) b_i\Bigr) 
  +H_{t}+H_J\nonumber\\ 
& &+\frac{i}{\sqrt{N}}\sum_{i}\lambda_i\Bigl(f^+_{i\sigma}
   f_{i\sigma}+(r_i+b_i^+)(r_i+b_i)-\frac{N}{2}\Bigr),
\nonumber\\
\\
&H&_t=-\frac{2t}{N}\sum_{<ij>}f^+_{i\sigma}f_{j\sigma}
   \bigl( b^+_jb_i+r_ir_j+r_jb_i+b_j^+r_i\bigr)+h.c.
\nonumber
\end{eqnarray}
Here the $\lambda$ field is introduced to enforce the constraint, and
$\mu_f,\mu_b$ are fixed by the particle number equations
$\langle n_f\rangle =\frac{N}{2} (1-\delta)$ and
$\langle r^2_i+b^+_ib_i\rangle =\frac{N}{2}\delta$, respectively.
The uniform mean field solution $r_i=r_0\sqrt{N/2}$ leads in the large
$N$ limit to the renormalized narrow fermionic spectrum 
$\xi_{\bf k}=-z\tilde t \gamma_{\bf k}-\mu_f$, with $\tilde t =J\chi+t\delta$, 
$\gamma_{\bf k}=\frac{1}{2}(\cos{k_x}+\cos{k_y})$,
$\chi=\sum_{\sigma}\langle f^+_{i\sigma} f_{j\sigma}\rangle /N$, and $z=4$ the
number of nearest neighbors. 
Distinct from the finite-temperature gauge-field theory of
Nagaosa and Lee\cite{nag90} the bond-order phase fluctuations acquire
a characteristic energy scale in this approach\cite{wan91}, and the
fermionic (``spinon'') excitations can be identified with Fermi-liquid
quasiparticles.
The mean field spectrum of bosons is
$\omega_{\bf q}=2z\chi t(1-\gamma_{\bf q})$. Thus the effective mass of holes
$m^0_h\propto 1/t$ is much smaller than that of the spinons. 

Due to the diluteness of the bosonic subsystem, $\delta\ll1$, the
density correlation function is mainly given by the condensate induced
part which is represented by the Green's function $\langle (b^+_{\bf q}
+b_{\bf -q})(b_{\bf q}+b^+_{\bf -q})\rangle_{\omega}$ 
for $q>q_0$, 
and $2\langle r_{\bf -q}r_{\bf q}\rangle_{\omega}$ for  $q<q_0$,
respectively.  
The $1/N$ self-energy corrections to these functions are calculated in
a conventional way\cite{rea83,kot88} expanding
$r_i=(r_0\sqrt{N}+(\delta r)_i)/\sqrt{2}$ and considering Gaussian
fluctuations  around the mean field solution. Neglecting all terms of
order $\delta/N$ and $q^2_0/N$, only one relevant $1/N$ contribution
remains which corresponds to the dressing of the slave-boson Green's
function by spinon particle-hole excitations. Within this
approximation and at zero temperature no divergences occur at low
momenta, thus one can take the limit $q_0\rightarrow0$. The final
result for the dynamic stucture factor (normalized by the hole
density) is: 
\begin{eqnarray}
N_{{\bf q},\omega}&=&\frac{2}{\pi} Im \Bigl( \bigl( \omega_{\bf q}a 
 +S^{(1/N)}_{{\bf q},\omega}-\mu_b\bigr)/D_{{\bf q},\omega} \Bigr),
\nonumber \\
\\
D_{{\bf q},\omega}&=& \bigl( \omega_{\bf q}a 
 +S^{(1/N)}_{{\bf q},\omega}-\mu_b\bigr) \bigl(\omega_{\bf q} 
 +S^{(1)}_{{\bf q},\omega}+S^{(1/N)}_{{\bf q},\omega}-\mu_b\bigr)
\nonumber\\
&-&\bigl(\omega a -A^{(1/N)}_{{\bf q},\omega}\bigr)^2.
\nonumber
\end{eqnarray}
The origin of the contribution
\begin{eqnarray}
S^{(1)}_{{\bf q},\omega}&=&ztr_0^2
\bigl(\frac{(1+\Pi_2)^2}{\Pi_1}-\Pi_3\bigr)_{{\bf q},\omega},
\\
\Pi_m&=&zt\sum_{\bf k}\frac{n(\xi_{\bf k})-n(\xi_{\bf k+q})}
 {\xi_{\bf k+q}-\xi_{\bf k}-\omega-i0^+}(\gamma_{\bf k}
 +\gamma_{\bf k+q})^{m-1},\nonumber
\end{eqnarray}
is the indirect interaction of bosons via the spinon band due to the
hopping term (which gives $\Pi_3$ in (4)) and due to the coupling to
spinons via the constraint field $\lambda$.
The latter channel provides a repulsion between bosons, making 
$S^{(1)}(\omega=0)$ positive and therefore ensuring the stability of
the uniform mean-field solution. The $1/N$ self energies $S^{(1/N)}$
and $A^{(1/N)}$ are essentially a single boson property. They are
given by the symmetric and antisymmetric combinations 
(with respect to $\omega+i0^+\rightarrow -\omega-i0^+$)
of the self energy
\begin{equation}
\Sigma^{(1/N)}_{{\bf q},\omega}=\frac{4}{N}
\sum_{|{\bf k}|<k_F<|{\bf k'}|} 
(zt\gamma_{\bf k'-q})^2 G^0_{\bf q+k-k'}  
(\omega+\xi_{\bf k}-\xi_{\bf k'}).
\end{equation}
Here $G^0_{\bf q}(\omega)=(\omega-\omega_{\bf q}-
\Sigma^{(1/N)}_{{\bf q},\omega}+\mu_b)^{-1}$ 
is the Green's function for a single slave boson moving in a 
uniform RVB background. 
Although in the context of $1/N$ theory the $G^0$ function in (5)
should be considered as a free propagator, we shall use here the
selfconsistent polaron picture for a single hole\cite{kan89}.
This is important when comparing the theory for $N=2$ with
diagonalization studies. Finally, the constants $a$ and $\mu_b$ in (3)
are given by $(1-t r^2_0/{\tilde t})$ and 
$S^{(1/N)}(\omega={\bf q}=0)$, respectively.

In the small $\omega,{\bf q}$ limit $N({\bf q},\omega)$ (3) is mainly
controlled by the interaction of bosons represented by the $S^{(1)}$
term, while the internal polaron structure of the boson determined by 
$S^{(1/N)}$ is less important even for $N=2$, as expected on physical
grounds. In this limit our results are essentially similar to those
obtained earlier\cite{wan91,geh95}.
$N({\bf q},\omega)$ consists of a weak spinon particle-hole continuum
with cutoff $\propto v_F q$, and a very pronounced collective mode
which nearly exhausts the sum rule. We find that the velocity of this
mode is always somewhat smaller than the Fermi velocity, $v_s\leq v_F
\simeq z\tilde t$, and therefore in a strict sense there is no well
defined sound. Since the spinon-boson coupling does not vanish in the
limit $q\rightarrow 0$ the imaginary part of the collective excitation
pole is linear in $\omega$ (or $q$), thus the damping is only
numerically small compared with the excitation energy.

The density response  $N({\bf q},\omega)$ at large momenta,
$q>\delta$, which we can compare with diagonalization results, is
dominated by the properties of a single boson dressed by spinon
excitations. 
For the numerical calculation of  $N({\bf q},\omega)$ given by 
Eqs. (3-5) we evaluate in a first step the single hole self-energy 
$\Sigma^{(1/N)}_{{\bf q},\omega}$ in a selfconsistent way. 
Then the parameter $r_0^2$ in (3-4), which formally corresponds to the
condensate fraction in our theory, is found selfconsistently from 
$r_0^2=\delta-\sum_{{\bf q}\neq 0}\tilde n_{\bf q}$. 

\begin{figure}
\leftline{
\psfig{width=8.5truecm,bbllx=60pt,bblly=55pt,bburx=580pt,bbury=700pt,angle=-90,file={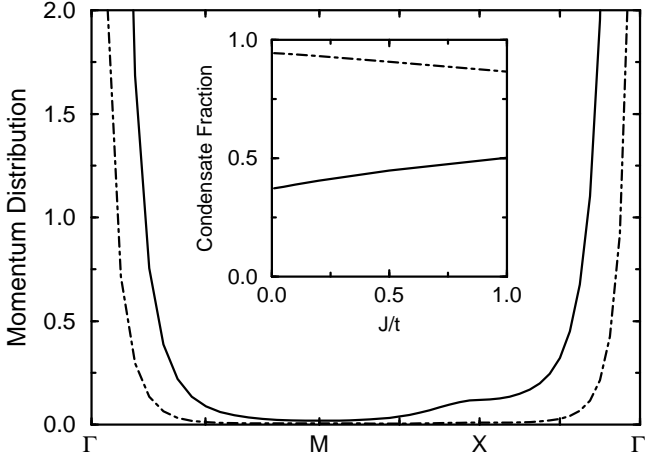}}}
\caption{Momentum distribution $\tilde n_{\bf q}$ normalized by
 $\delta$, for $\delta=0.25$ and $J/t=0.4$. Inset: Condensate fraction
 $r_0^2/\delta$ versus J for $\delta=0.25$. Comparison of calculation
 with (solid line) and without (dash-dotted) inclusion of polaron
 effects.}
\label{fig:one}
\end{figure}

The momentum
distribution $\tilde n_{\bf q}=\langle b^+_{\bf q}b_{\bf q}\rangle$ is
calculated using the corresponding bosonic Green's function for finite
hole-density:
\begin{eqnarray}
G_{{\bf q},\omega}=-\Bigl(&\omega&a-A^{(1/N)}_{{\bf q},\omega} 
 +\omega_{\bf q}\bigl(1-\frac{t r_0^2}{2\tilde t}\bigr) 
\nonumber\\
 &+&\frac{1}{2}S^{(1)}_{{\bf q},\omega}+S^{(1/N)}_{{\bf q},\omega}
 -\mu_b \Bigr)/D_{{\bf q},\omega}.
\end{eqnarray}
While the $1/q$ behavior of $\tilde n_{\bf q}$ (Fig.1) at small
momenta, $q\leq\delta$, 
is provided by the presence of the condensate, the momentum
distribution in the rest of the Brillouin zone is determined by the
polaron effect , i.e. by the  renormalization of the slave boson wave
function due to the strong local scattering from spinons. The latter
leads to a strong reduction of the condensate fraction $r_0^2$, as
shown in the inset of Fig.1.
The $\omega$-dependence of the dynamic structure factor $N({\bf
  q},\omega)$ given by Eq. (3) is shown in Fig. 2 for momenta along
the direction $(\pi,0)$. Results for some selected momenta are
presented in Fig.3 together with the exact diagonalization data\cite{toh95}. 
We note that the overall energy scale for $N({\bf q},\omega)$   at
large momenta is insensitive to the ratio $J/t$, whereas the value of
the spinon bond-order parameter $\chi$, which enters in (3) via the 
free bosonic dispersion $\omega_{\bf q}$, is important. 
In the $N=\infty $ limit $\chi_{\infty}\simeq 2/\pi^2$ is given by that of
free fermions, while for the original $t$-$J$ model its value should
be larger\cite{zha88} due to Gutzwiller projection.
For our comparison of  $N({\bf q},\omega)$ (Eq. (3)) with the exact
results of Ref.\cite{toh95} we consider $\chi$ as a free parameter
chosing $\chi=\frac{3}{2}\chi_{\infty}$. 

\begin{figure}
\vspace{-35mm}
\centerline{
\psfig{width=8.5truecm,bbllx=60pt,bblly=55pt,bburx=580pt,bbury=500pt,angle=-90,file={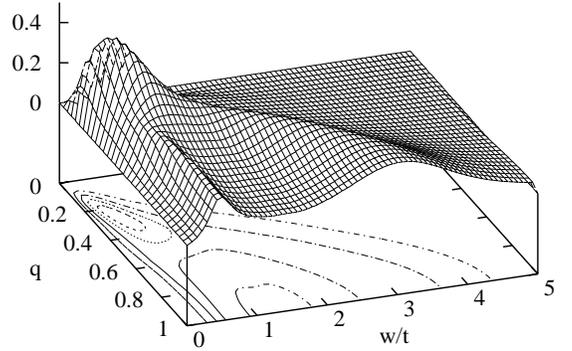}}}
\caption{Density response function $N({\bf q},\omega)$ for $J/t=0.4$
 and $\delta=0.25$ along the $(\pi,0)$ direction. }
\label{fig:two}
\end{figure}

The calculated density response function of the $t$-$J$ model has
quite rich structure showing pronounced features on different energy
scales.
The main spectral weight of the excitations at large momenta is
located in an energy region of order of several $t$. This high energy
peak is very broad and incoherent as a result of the strong coupling
of bosons to low-energy spin excitations. We find that the position of
this peak and its shape are rather insensitive to the ratio $J/t\leq
1$ in agreement with conclusions of \cite{toh95,ede95}.
This is simply due to the fact that the high-energy properties of the
$t$-$J$ model are controlled by $t$.
The theory predicts also a second peak at lower energy (Figs. 2,3)
which is more pronounced in the direction $(\pi,0)$, while its weight
is strongly suppressed for ${\bf q}$ near $(\pi,\pi)$. 
The origin of this excitation is due to the formation of a
polaron-like band of dressed bosons. The relative weight of this
contribution  increases with $J$ as a result of the increasing
spinon bandwidth. As noted above, the density response for small
$q$ is mainly given by the collective excitation with energy
$\epsilon_q =v_s q$. The velocity $v_s$ is an increasing function of the
hole density $\delta$ as expected.
Distinct from the one-dimensional model here the collective density 
(``holon'') excitations always overlap with the spinon particle-hole
spectrum leading to a strong damping which is also linear in $q$,
i.e. $\gamma_q=\alpha \epsilon_q$. 
Numerically, at $\delta=0.25$ and $J=0.4 t$ for instance, we found
$v_s\simeq 0.4 t$ and $\alpha \simeq 0.5$.

\begin{figure}
\leftline{
\psfig{width=8.5truecm,bbllx=60pt,bblly=55pt,bburx=580pt,bbury=580pt,angle=-90,file={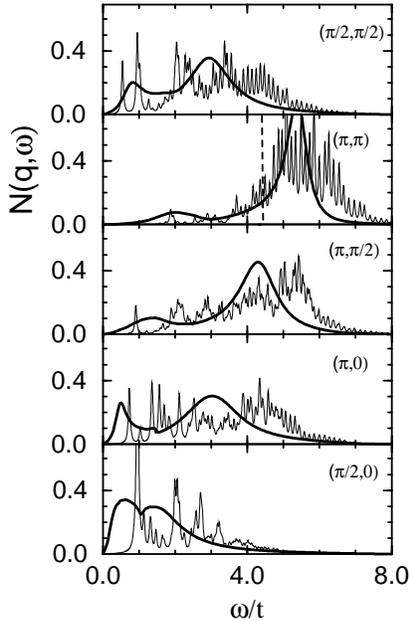}}}
\caption{Comparison of $N({\bf q},\omega)$ for large momenta with data
 obtained by exact diagonalization\protect{\cite{toh95}} 
 for a $4\times4$
 periodic cluster with $J/t=0.4$ and $\delta=0.25$. 
 The dashed line in the $(\pi,\pi)$ spectrum indicates the
 $\delta$-function peak obtained if polaron effects are neglected,
 i.e. if $\Sigma^{(1/N)}=0$ . }
\label{fig:three}
\end{figure}

Finally we discuss our results for the static structure factor 
 $N({\bf q})=\delta\cdot \int d\omega N({\bf q},\omega)$, where the 
factor $\delta$ appears because $ N({\bf q},\omega)$ denotes the
normalized density response. Figure 4 shows how
 effectively the theory presented accounts for the strong suppression
 of density fluctuations in the vicinity of the Mott transition.
The calculated $N({\bf q})$ is a linear function of $q$ for small
 momenta as a result of the interaction between bosons, and it
 saturates at a value $N({\bf q})\simeq \delta$ for large $q$ as it
 should. Our result for $N({\bf q})$ is very close to that by Gros and
 Valenti\cite{gro94}, who used the Gutzwiller projected wave function for
 constrained fermions. It is worth noticing that our results fulfil
 the sum rule $\sum N({\bf q})=\delta(1-\delta)$ for constrained
 electrons within a few percent accuracy.

In conclusion we have studied the density fluctuation spectrum of the
$t$-$J$ model at moderate doping and zero temperature. 
Our theory based on the assumption
of a Fermi-liquid ground state of this model captures all essential
features observed in exact diagonalization studies\cite{toh95,ede95,pre96}. 
Concerning the high-$T_c$ materials we predict in the density response
at large momentum transfer a low energy peak with energy of the order
of $z\tilde t=z (J\chi+t\delta )$ 
arising from the coherent polaron motion of
holes, while the linear collective mode will be changed into a plasmon
mode due to the Coulomb interaction. The latter has been already  
investigated by electron-energy loss techniques\cite{nue89}, however
the scale  $z\tilde t$ required to see
the low energy structure in $N({\bf q},\omega)$ has not yet been
reached in these experiments.

\begin{figure}
\leftline{
\psfig{width=8.0truecm,bbllx=60pt,bblly=55pt,bburx=580pt,bbury=700pt,angle=-90,file={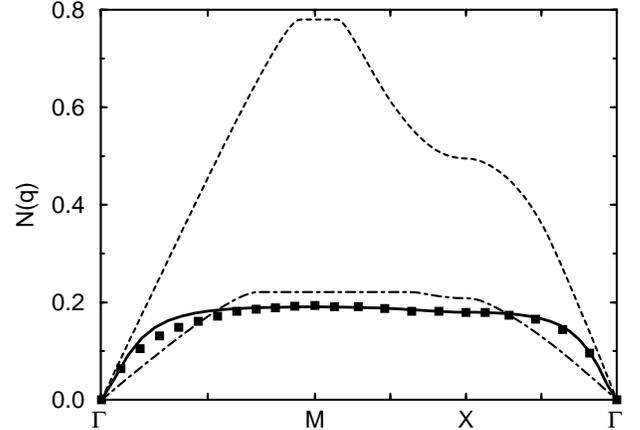}}}
\caption{Comparison of static structure factor $N(\bf{q})$ (solid line)
 calculated for $J=0.4$ with the result 
 for the Gutzwiller wave function of 
 Gros and Valenti\protect{\cite{gro94}}
 (squares) for a hole concentration $\delta=0.213$.
 For comparison the results for spinless fermions (dash-dotted
 line) and fermions with spin (dashed line) are shown.  
 The latter comparison shows
 the large reduction of the density response due to the constraint. }
\label{fig:four}
\end{figure}

We acknowledge useful discussions with P. Prelov\v sek, T. Tohyama 
and R. Zeyher,
and thank C. Gros for providing their Gutzwiller data. One of us (GKh)
would like to thank L. Hedin and the Max-Planck-Institut FKF for the
hospitality extended to him during his stay.
This work was supported in part by E.U. Grant no ERBCHRX CT94.0438.

\narrowtext


\end{document}